\documentclass[floatfix,aps,showpacs,twocolumn,nofootinbib,preprintnumbers]
{revtex4}
\usepackage{graphicx}

\begin{document}

\preprint{FERMILAB-PUB-06-106-A, CERN-PH-TH/2006-087}

\title{Inflation model constraints from the\\ 
Wilkinson Microwave Anisotropy Probe
three-year data}
\author{William H.\ Kinney} \email{whkinney@buffalo.edu}
\affiliation{Department of Physics, University at Buffalo,
        the State University of New York, Buffalo, NY 14260-1500}
\author{Edward W.\ Kolb} \email{rocky@fnal.gov}
\affiliation{Particle Astrophysics Center, Fermi
       	National Accelerator Laboratory, Batavia, Illinois \ 60510-0500, USA,\\
       	and Department of Astronomy and Astrophysics, Enrico Fermi Institute,
       	University of Chicago, Chicago, Illinois \ 60637-1433, USA}
\author{Alessandro Melchiorri} \email{alessandro.melchiorri@roma1.infn.it}
\affiliation{Dipartimento di Fisica and Sezione INFN,
Universita' di Roma ``La Sapienza'', Ple Aldo Moro 2, 00185, Italy}
\author{Antonio Riotto} \email{antonio.riotto@pd.infn.it}
\affiliation{CERN, Theory Division, Geneva 23, CH-1211, Switzerland}

\date{\today}

\begin{abstract}

We extract parameters relevant for distinguishing among single-field inflation
models from the Wilkinson Microwave Anisotropy Probe (WMAP) three-year data
set, and also from WMAP in combination with the Sloan Digital Sky Survey (SDSS)
galaxy power spectrum. Our analysis leads to the following conclusions: 1) the
Harrison--Zel'dovich model is consistent with both data sets at a 95\%
confidence level; 2) there is no strong evidence for running of the spectral
index of scalar perturbations; 3) Potentials of the form $V \propto \phi^p$ are
consistent with the data for $p = 2$, and are marginally consistent with the
WMAP data considered alone for $p = 4$, but ruled out by WMAP combined with
SDSS. We perform a ``Monte Carlo reconstruction'' of the inflationary
potential, and find that: 1) there is no evidence to support an observational
lower bound on the amplitude of gravitational waves produced during inflation;
2) models such as simple hybrid potentials which evolve toward an inflationary
late-time attractor in the space of flow parameters are strongly disfavored by
the data, 3) models selected with even a weak slow-roll prior strongly cluster
in the region favoring a ``red'' power spectrum and no running of the spectral
index, consistent with simple single-field inflation models.

\end{abstract}


\maketitle

%%%%%%%%%%%%%%%%%%%%%%%%%%%%%%%%%%%%%%%%%%%%%%%%%%%%%%%%%%%%%%%%%%%%%%%%%
%%%%%%%%%%%%%%%%%%%%%%%%%%%%%%%%%%%%%%%%%%%%%%%%%%%%%%%%%%%%%%%%%%%%%%%%%
\section{Introduction}
%%%%%%%%%%%%%%%%%%%%%%%%%%%%%%%%%%%%%%%%%%%%%%%%%%%%%%%%%%%%%%%%%%%%%%%%%
%%%%%%%%%%%%%%%%%%%%%%%%%%%%%%%%%%%%%%%%%%%%%%%%%%%%%%%%%%%%%%%%%%%%%%%%%

Inflation \cite{lrreview} has become the dominant paradigm for understanding
the initial conditions for structure formation and for Cosmic Microwave
Background (CMB) temperature anisotropies. In the inflationary picture,
primordial density and gravitational-wave fluctuations are created from quantum
fluctuations, ``redshifted'' beyond the horizon during an early period of
superluminal expansion of the universe, then ``frozen''
\cite{Starobinsky:1979ty,muk81,bardeen83}.  Perturbations at the surface of
last scattering are observable as temperature anisotropies in the CMB, as first
detected by the Cosmic Background Explorer  satellite
\cite{bennett96,gorski96}. The latest and most impressive confirmation of the
inflationary paradigm has been recently provided by the three-year data from
the Wilkinson Microwave Anisotropy Probe (WMAP) satellite
\cite{wmap3cosm,wmap3pol,wmap3temp,wmap3beam}.  The WMAP collaboration has
produced new full-sky temperature maps in five frequency bands from 23 to 94
GHz based on the first three years of the WMAP sky survey. The new maps, which
are consistent with the first-year maps and more sensitive, incorporate
improvements in data processing made possible by the additional years of data
and by a more complete analysis of the polarization signal.  WMAP data support
the inflationary model as the mechanism for the generation of super-horizon
curvature fluctuations.

The goal of this paper is to make use of the recent WMAP three-year data
(WMAP3) to discriminate among the various single-field inflationary models. As
such, this paper represents a complete update of our previous analysis
\cite{wmapping1} of the first-year WMAP data.   

For single-field inflation models, the relevant parameter space for
distinguishing among models is defined by the scalar spectral index $n$, the
ratio of tensor to scalar fluctuations $r$, and the running of the scalar
spectral index $d n / d\ln{k}$.  We employ {\em Monte Carlo reconstruction}, a
stochastic method for ``inverting'' observational constraints to generate an
ensemble of inflationary potentials compatible with observation
\cite{kinney02,easther02}.  In addition to encompassing a broader set of models
than usually considered (large-field, small-field, hybrid and linear models),
Monte Carlo reconstruction makes it possible easily to include effects to
higher order in slow roll. 

The paper is organized as follows: In Sec.\ II we will quickly review
single-field inflation models and their observables. In Sec.\ III we define the
inflationary model space as a function of the slow-roll parameters $\epsilon$
and $\eta$. In Sec.\ IV we describe our analysis method as well as our results.
Since a study of the implications of the WMAP3 data for single field models of
inflation has been already performed by the WMAP collaboration themselves
\cite{wmap3cosm}, we will also specify briefly the differences between our
analysis and theirs.  In Sec.\ V we describe a Monte Carlo reconstruction
method to determine an ensemble of inflationary potentials compatible with
observations.  In Sec.\ VI we present our conclusions.

%%%%%%%%%%%%%%%%%%%%%%%%%%%%%%%%%%%%%%%%%%%%%%%%%%%%%%%%%%%%%%%%%%%%%%%%%
%%%%%%%%%%%%%%%%%%%%%%%%%%%%%%%%%%%%%%%%%%%%%%%%%%%%%%%%%%%%%%%%%%%%%%%%%
\section{Single-field inflation and the inflationary observables}
%%%%%%%%%%%%%%%%%%%%%%%%%%%%%%%%%%%%%%%%%%%%%%%%%%%%%%%%%%%%%%%%%%%%%%%%%
%%%%%%%%%%%%%%%%%%%%%%%%%%%%%%%%%%%%%%%%%%%%%%%%%%%%%%%%%%%%%%%%%%%%%%%%%

In this section we briefly review scalar field models of inflationary
cosmology, and explain how we relate model parameters to observable quantities.
Inflation, in its most general sense, can be defined to be a period of
accelerating cosmological expansion during which the universe evolves toward
homogeneity and flatness.  Within this broad framework, many specific models
for inflation have been proposed. We limit ourselves here to models with
``normal'' gravity ({\em i.e.,} general relativity) and a single order
parameter for the vacuum, described by a slowly rolling scalar field $\phi$,
the inflaton. 

A scalar field in a cosmological background evolves with an equation of motion
$\ddot\phi + 3 H \dot\phi + V'\left(\phi\right) = 0.$ The evolution of the
scale factor is given by the scalar-field dominated FRW equation,
\begin{eqnarray}
H^2 & &= {8 \pi \over 3 m_{\rm Pl}^2} \left[{1 \over 2} \dot\phi^2 +
V\left(\phi\right)\right],\cr
\left(\ddot a \over a\right) &&= {8 \pi \over 3 m_{\rm Pl}^2}
\left[V\left(\phi\right) - \dot\phi^2\right].
\label{eqbackground}
\end{eqnarray}
We have assumed a flat Friedmann-Robertson-Walker metric $g_{\mu \nu} = {\rm
diag}(1, -a^2, -a^2 -a^2)$, where $a^2(t)$ is the scale factor of the universe.
{\em Inflation} is defined to be a period of accelerated expansion, $\ddot a >
0$.  A powerful way of describing the dynamics of a scalar field-dominated
cosmology is to express the Hubble parameter as a function of the field $\phi$,
$H = H(\phi)$, which is consistent provided $\phi$ is monotonic in time. The
equations of motion become \cite{grishchuk88,muslimov90,salopek90,lidsey95}:
\begin{eqnarray} 
& &\dot\phi =  -{m_{\rm Pl}^2 \over 4 \pi} H'(\phi),\cr
& & \left[H'(\phi)\right]^2 - {12 \pi \over m_{\rm Pl}^2}
H^2(\phi) = - {32 \pi^2 \over m_{\rm Pl}^4}
V(\phi).
\label{eqbasichjequations}
\end{eqnarray}
These are completely equivalent to the second-order equation of motion. The
second of the above equations is referred to as the {\it Hamilton-Jacobi}
equation, and can be written in the useful form
\begin{equation} 
H^2(\phi) \left[1 - {1\over 3}
\epsilon(\phi)\right] =  \left({8 \pi \over 3 m_{\rm Pl}^2}\right) V(\phi),
\label{eqhubblehamiltonjacobi}
\end{equation}
where $\epsilon$ is defined to be
\begin{equation}
\epsilon(\phi) \equiv {m_{\rm Pl}^2 \over 4 \pi} \left({H'(\phi) \over
 H(\phi)}\right)^2.\label{eqdefofepsilon}
\end{equation}
The physical meaning of $\epsilon(\phi)$ can be seen by expressing Eq.\
(\ref{eqbackground}) as
\begin{equation}
\left({\ddot a \over a}\right) = H^2 (\phi) \left[1 -
 \epsilon(\phi)\right],
\end{equation}
so that the condition for inflation, $(\ddot a / a) > 0$, is equivalent to
$\epsilon < 1$. The scale factor is given by
\begin{equation}
a \propto e^{N} = \exp\left[\int_{t_0}^{t}{H\,dt}\right],
\end{equation}
where the number of {\it e}-folds $N$ is
\begin{equation}
N \equiv \int_{t}^{t_e}{H\,dt} = \int_{\phi}^{\phi_e}{{H \over
\dot\phi}\,d\phi} = {2 \sqrt{\pi} \over m_{\rm Pl}}
\int_{\phi_e}^{\phi}{d\phi \over
\sqrt{\epsilon(\phi)}}.\label{eqdefofN}
\end{equation}

We will frequently work within the context of the {\em slow-roll}
approximation, which is the assumption that the evolution of the field is
dominated by the drag from the cosmological expansion, so that $\ddot\phi
\simeq 0$ and  $\dot \phi \simeq -V'/3 H$. The equation of state of the scalar
field is dominated by the potential, so that $p \simeq -\rho$, and the
expansion rate is approximately $H^2 \simeq 8 \pi V(\phi)/ 3 m_{\rm Pl}^2$. 
The slow roll approximation is consistent if both the slope and curvature of
the potential are small, $V',\ V'' \ll V$. In this case the parameter
$\epsilon$ can be expressed in terms of the potential as
\begin{equation}
\epsilon \equiv {m_{\rm Pl}^2 \over 4 \pi} \left({H'\left(\phi\right) \over
H\left(\phi\right)}\right)^2 \simeq {m_{\rm Pl}^2 \over 16 \pi}
\left({V'\left(\phi\right) \over V\left(\phi\right)}\right)^2.
\end{equation}
We will also define a second ``slow-roll parameter'' $\eta$ by
\begin{eqnarray}
\eta\left(\phi\right) &\equiv& {m_{\rm Pl}^2 \over 4 \pi} 
\left({H''\left(\phi\right)
\over H\left(\phi\right)}\right)\cr
&\simeq& {m_{\rm Pl}^2 \over 8 \pi}
\left[{V''\left(\phi\right) \over V\left(\phi\right)} - {1 \over 2}
\left({V'\left(\phi\right) \over V\left(\phi\right)}\right)^2\right].
\end{eqnarray}
Slow roll is then a consistent approximation for $\epsilon,\ \eta \ll 1$. 

Inflation models not only explain the large-scale homogeneity of the universe,
but also provide a mechanism for explaining the observed level of {\em
inhomogeneity} as well. During inflation, quantum fluctuations on small scales
are quickly redshifted to scales much larger than the horizon size, where they
are ``frozen'' as perturbations in the background metric. The metric
perturbations created during inflation are of two types: scalar, or {\it
curvature} perturbations, which couple to the stress-energy of matter in the
universe and form the ``seeds'' for structure formation, and tensor, or
gravitational-wave perturbations, which do not couple to matter.  Both scalar
and tensor perturbations contribute to CMB anisotropy. Scalar fluctuations can
also be interpreted as fluctuations in the density of the matter in the
universe. Scalar fluctuations can be quantitatively characterized by the
comoving curvature perturbation $P_{\cal R}$. As long as the equation of state
$\epsilon$ is slowly varying, the curvature perturbation can be shown to be
\cite{lrreview}
\begin{equation}
P_{\cal R}^{1/2}\left(k\right) = \left({H^2 \over 2 \pi \dot \phi}\right)_{k =
a H} =    \left [{H \over m_{\rm Pl} } {1 \over \sqrt{\pi \epsilon}}\right]_{k
= a H}.
\end{equation}
The fluctuation power spectrum is in general a function of wavenumber $k$, and
is evaluated when a given mode crosses outside the horizon during inflation, $k
= a H$. Outside the horizon, modes do not evolve, so the amplitude of the mode
when it crosses back {\em inside} the horizon during a later radiation- or
matter-dominated epoch is just its value when it left the horizon during
inflation.  Instead of specifying the fluctuation amplitude directly as a
function of $k$, it is convenient to specify it as a function of the number of
{\it e}-folds $N$ before the end of inflation at which a mode crossed outside
the horizon. 

The {\em spectral index} $n$ for $P_{\cal R}$ is defined by
\begin{equation}
n - 1 \equiv {d\ln P_{\cal R} \over d\ln k},
\end{equation}
so that a scale-invariant spectrum, in which modes have constant amplitude at
horizon crossing, is characterized by $n = 1$. 

The power spectrum of tensor fluctuation modes is given
by \cite{lrreview}
\begin{equation}
P_{T}^{1/2}\left(k_N\right) = \left[\frac{4 H}{m_{\rm Pl} \sqrt{\pi}}
\right]_{N}.
\end{equation}
The ratio of tensor-to-scalar modes is then $ P_{T}/P_{\cal R} = 16 \epsilon$,
so that tensor modes are negligible for $\epsilon \ll 1$.\footnote{This
normalization convention is different from that used in our analysis of the
first-year WMAP data \cite{wmapping1}, which used the convention $r = 10
\epsilon$. In this paper, we have adopted the more standard normalization
convention $r = 16
\epsilon$.}

%%%%%%%%%%%%%%%%%%%%%%%%%%%%%%%%%%%%%%%%%%%%%%%%%%%%%%%%%%%%%%%%%%%%%%%%%
%%%%%%%%%%%%%%%%%%%%%%%%%%%%%%%%%%%%%%%%%%%%%%%%%%%%%%%%%%%%%%%%%%%%%%%%%
\section{The inflationary model space}
\label{seczoology}
%%%%%%%%%%%%%%%%%%%%%%%%%%%%%%%%%%%%%%%%%%%%%%%%%%%%%%%%%%%%%%%%%%%%%%%%%
%%%%%%%%%%%%%%%%%%%%%%%%%%%%%%%%%%%%%%%%%%%%%%%%%%%%%%%%%%%%%%%%%%%%%%%%%

To summarize the results of the previous section, inflation generates scalar
(density) and tensor (gravitational-wave) fluctuations which are generally well
approximated by power laws: 
$P_{\cal R}\left(k\right) \propto k^{n - 1}$, $P_{T}\left(k\right) 
\propto k^{n_{T}}$. 
In the limit of slow roll, the spectral indices $n$ and $n_{T}$ vary slowly or
not at all with scale.  We can write the spectral indices $n$ and $n_{T}$ to
lowest order in terms of the slow roll parameters $\epsilon$ and $\eta$ as:
\begin{eqnarray}
n \simeq&& 1 - 4 \epsilon + 2 \eta,\cr
n_{T} \simeq&& - 2 \epsilon.
\end{eqnarray}
The tensor/scalar ratio is frequently expressed as a quantity $r$, which is 
conventionally normalized as
\begin{equation}
r \equiv 16 \epsilon = {P_{\rm T} \over P_{\cal R}} .
\end{equation}
The tensor spectral index is {\em not} an independent parameter, but is
proportional to the tensor/scalar ratio, given to lowest order in slow roll by
$ n_{T} \simeq - 2 \epsilon = - r/8$.  This is known as the consistency
relation for inflation.  A given inflation model can therefore be described to
lowest order in slow roll by three independent parameters, $P_{\cal R}$,
$P_{T}$, and $n$. If we wish to include higher-order effects, we have a fourth
parameter describing the running of the scalar spectral index, $d n / d\ln{k}$.

Calculating the CMB fluctuations from a particular inflation model reduces
to the following basic steps: (1) from the potential, calculate $\epsilon$ and
$\eta$. (2) From $\epsilon$, calculate $N$ as a function of the field $\phi$.
(3) Invert $N\left(\phi\right)$ to find $\phi_N$. (4) Calculate $P_{\cal R}$,
$n$, and $P_T$ as functions of $\phi$, and evaluate them at $\phi =
\phi_N$. For the remainder of the paper, all parameters are assumed to be
evaluated at $\phi = \phi_N$, where $N$ varies from $46$ to $60$. 

With the normalization fixed, the relevant parameter space for distinguishing
between inflation models to lowest order in slow roll is then the $r$---$n$
plane.  (To next order in slow-roll parameters, one must introduce the running
of $n$.)  Different classes of models are distinguished by the value of the
second derivative of the potential, or, equivalently, by the relationship
between the values of the slow-roll parameters $\epsilon$ and $\eta$.  Each
class of models has a different relationship between $r$ and $n$. For a more
detailed discussion of these relations, the reader is referred to Refs.\
\cite{dodelson97,kinney98a}.  

Even restricting ourselves to a simple single-field inflation scenario, the
number of models available to choose from is large \cite{lrreview}.  It is
convenient to define a general classification scheme, or ``zoology'' for models
of inflation. We divide models into three general types: {\it large-field},
{\it small-field}, and {\it hybrid}, with a fourth classification, {\it linear}
models, serving as a boundary between large- and small-field models. 

First order in $\epsilon$ and $\eta$ is sufficiently accurate for the purposes
of this Section, and for the remainder of this Section we will only work to
first order. The generalization to higher order in slow roll will be discussed 
in the following.

%%%%%%%%%%%%%%%%%%%%%%%%%%%%%%%%%%%%%%%%%%%%%%%%%%%%%%%%%%%%%%%%%%%%%%%%%
%%%%%%%%%%%%%%%%%%%%%%%%%%%%%%%%%%%%%%%%%%%%%%%%%%%%%%%%%%%%%%%%%%%%%%%%%
\subsection{Large-field models: $-\epsilon < \eta \leq \epsilon$}
%%%%%%%%%%%%%%%%%%%%%%%%%%%%%%%%%%%%%%%%%%%%%%%%%%%%%%%%%%%%%%%%%%%%%%%%%
%%%%%%%%%%%%%%%%%%%%%%%%%%%%%%%%%%%%%%%%%%%%%%%%%%%%%%%%%%%%%%%%%%%%%%%%%

Large-field models have inflaton potentials typical of ``chaotic'' inflation
scenarios \cite{linde83}, in which the scalar field is displaced from the
minimum of the potential by an amount usually of order the Planck mass. Such
models are characterized by $V''\left(\phi\right) > 0$, and $-\epsilon < \eta
\leq \epsilon$. The generic large-field potentials we consider are polynomial
potentials $V\left(\phi\right) = \Lambda^4 \left({\phi / \mu}\right)^p$,
and exponential potentials, $V\left(\phi\right) = \Lambda^4 \exp\left({\phi /
\mu}\right)$. 

For the case of an exponential potential, $V\left(\phi\right)
\propto \exp\left({\phi / \mu}\right)$, the tensor/scalar ratio $r$ is simply
related to the spectral index as
\begin{equation}
r = 8 \left(1 - n\right),
\end{equation}
but the slow roll parameters have no dependence on the number of e-folds $N$. 

For inflation with a polynomial potential, $V\left(\phi\right) \propto \phi^p$,
we have
\begin{eqnarray}
n-1&=&-\frac{2+p}{2N}\, ,\nonumber\\
r&=&\frac{8p}{2N}=8 \left({p \over p + 2}\right) \left(1 - n\right)\, ,
\end{eqnarray}
so that tensor modes are large for significantly tilted spectra.

%%%%%%%%%%%%%%%%%%%%%%%%%%%%%%%%%%%%%%%%%%%%%%%%%%%%%%%%%%%%%%%%%%%%%%%%%
%%%%%%%%%%%%%%%%%%%%%%%%%%%%%%%%%%%%%%%%%%%%%%%%%%%%%%%%%%%%%%%%%%%%%%%%%
\subsection{Small-field models: $\eta < -\epsilon$}
%%%%%%%%%%%%%%%%%%%%%%%%%%%%%%%%%%%%%%%%%%%%%%%%%%%%%%%%%%%%%%%%%%%%%%%%%
%%%%%%%%%%%%%%%%%%%%%%%%%%%%%%%%%%%%%%%%%%%%%%%%%%%%%%%%%%%%%%%%%%%%%%%%%

Small-field models are the type of potentials that arise naturally from
spontaneous symmetry breaking (such as the original models of ``new'' inflation
\cite{linde82,albrecht82}) and from pseudo Nambu-Goldstone modes (natural
inflation \cite{freese90}). The field starts from near an unstable equilibrium
(taken to be at the origin) and rolls down the potential to a stable minimum.
Small-field models are typically characterized by $V''\left(\phi\right) < 0$
and $\eta < -\epsilon$. Typically $\epsilon$ (and hence the tensor amplitude)
is close to zero in small-field models. The generic small-field potentials we
consider are of the form $V\left(\phi\right) = \Lambda^4 \left[1 - \left({\phi
/ \mu}\right)^p\right]$, which can be viewed as a lowest-order Taylor expansion
of an arbitrary potential about the origin. The cases $p = 2$ and $p > 2$ have
very different behavior. For $p = 2$, $n-1\simeq -(1/2\pi)(m_{\rm Pl}/\mu)^2$
and there is no dependence upon the number of {\it e}-foldings. On the other
hand 
\begin{equation}
r = 8 (1 - n) \exp\left[- 1 - N\left(1 - n\right)\right].
\end{equation}

For $p > 2$, the scalar spectral index is
\begin{equation}
n \simeq 1 - {2 \over N} \left({p - 1 \over p - 2}\right),
\end{equation}
{\it independent} of $(m_{\rm Pl}/\mu)$. Assuming $\mu < m_{\rm Pl}$ results in
an upper bound on $r$ of
\begin{equation}
r < 8 {p \over N \left(p - 2\right)} \left[{8 \pi \over N p \left(p -
2\right)}\right]^{p / \left(p - 2\right)}.
\end{equation}

%%%%%%%%%%%%%%%%%%%%%%%%%%%%%%%%%%%%%%%%%%%%%%%%%%%%%%%%%%%%%%%%%%%%%%%%%
%%%%%%%%%%%%%%%%%%%%%%%%%%%%%%%%%%%%%%%%%%%%%%%%%%%%%%%%%%%%%%%%%%%%%%%%%
\subsection{Hybrid models: $0 < \epsilon < \eta$}
%%%%%%%%%%%%%%%%%%%%%%%%%%%%%%%%%%%%%%%%%%%%%%%%%%%%%%%%%%%%%%%%%%%%%%%%%
%%%%%%%%%%%%%%%%%%%%%%%%%%%%%%%%%%%%%%%%%%%%%%%%%%%%%%%%%%%%%%%%%%%%%%%%%

The hybrid scenario \cite{linde91,linde94,copeland94,lr97} frequently appears
in models which incorporate supersymmetry into inflation. In a typical
hybrid-inflation model, the scalar field responsible for inflation evolves
toward a minimum with nonzero vacuum energy. The end of inflation arises as a
result of instability in a second field. Such models are characterized by
$V''\left(\phi\right) > 0$ and $0 < \epsilon < \eta$. We consider generic
potentials for hybrid inflation of the form $V\left(\phi\right) = \Lambda^4
\left[1 + \left({\phi / \mu}\right)^p\right].$ The field value at the end of
inflation is determined by some other physics, so there is a second free
parameter characterizing the models. Because of this extra freedom, hybrid
models fill a broad region in the $r$---$n$ plane. For $\left({\phi_N /
\mu}\right)\gg 1$ (where $\phi_N$ is the value of the inflaton field when there
are $N$ {\it e}-foldings till the end of inflation) one recovers the same
results as the large-field models. On the contrary, when $\left({\phi_N /
\mu}\right)\ll 1$, the dynamics are analogous to small-field models, except
that the field is evolving toward, rather than away from, a dynamical fixed
point. While in principle ``hybrid'' models can populate a broad region of the
inflationary parameter space, the presence of a dynamical fixed point means
that there is a simple subclass of hybrid models that live in a narrow band of
parameter space along a line with $r \simeq 0$, $n > 1$, and $dn/d\ln{k} \simeq
0$.  We will see below that while the WMAP3 data do not rule out the entire
region which we label here as ``hybrid,'' the simplest hybrid models evolving
near the dynamical fixed point are clearly disfavored by the data.

An example of a model which falls into the``hybrid'' region of the $r$---$n$
plane away from the dynamical fixed point is a potential with a negative power 
of the scalar field, $V\left(\phi\right) =V_0\left[ 1+\alpha\left(m_{\rm
Pl}/\phi\right)^p\right]$, used in intermediate inflation \cite{barrow93} and
dynamical supersymmetric inflation \cite{kinney97}. The power spectrum is blue:
the spectral index given by $n-1\simeq 2(p+1)/[(p+2)(N_{\rm tot}-N)]$, where
$N_{\rm tot}$ is the total number of {\it e}-foldings, and the parameter $r$ is
generally negligible. However, the model exhibits running of the spectral index
which would be potentially detectable by future experiments,
\begin{equation}
\label{eq:dsirunning}
{dn \over d\ln{k}} = -{1 \over 2} \left({p + 2 \over p + 1}\right) 
\left(n - 1\right)^2.
\end{equation}
For example, for $p = 2$ and $n = 1.2$, the running is $dn / d\ln{k} =
 -0.05$ \cite{kinney98}. When the running is sizable, the  
tensor contribution is totally  negligible, 
\begin{equation}
r\ll P_{\cal  R}^{1/2}(n-1)^{(3p+5)/(p+2)}.
\end{equation}

%%%%%%%%%%%%%%%%%%%%%%%%%%%%%%%%%%%%%%%%%%%%%%%%%%%%%%%%%%%%%%%%%%%%%%%%%
%%%%%%%%%%%%%%%%%%%%%%%%%%%%%%%%%%%%%%%%%%%%%%%%%%%%%%%%%%%%%%%%%%%%%%%%%
\subsection{Linear models: $\eta = - \epsilon$}
%%%%%%%%%%%%%%%%%%%%%%%%%%%%%%%%%%%%%%%%%%%%%%%%%%%%%%%%%%%%%%%%%%%%%%%%%
%%%%%%%%%%%%%%%%%%%%%%%%%%%%%%%%%%%%%%%%%%%%%%%%%%%%%%%%%%%%%%%%%%%%%%%%%

Linear models, $V\left(\phi\right) \propto \phi$, live on the boundary between
large-field and small-field models, with $V''\left(\phi\right) = 0$ and $\eta =
- \epsilon$. The spectral index and tensor/scalar ratio are related as
\begin{equation}
r = {8 \over 3} \left(1 - n\right).
\end{equation}

%%%%%%%%%%%%%%%%%%%%%%%%%%%%%%%%%%%%%%%%%%%%%%%%%%%%%%%%%%%%%%%%%%%%%%%%%
%%%%%%%%%%%%%%%%%%%%%%%%%%%%%%%%%%%%%%%%%%%%%%%%%%%%%%%%%%%%%%%%%%%%%%%%%
\subsection{Other models}
%%%%%%%%%%%%%%%%%%%%%%%%%%%%%%%%%%%%%%%%%%%%%%%%%%%%%%%%%%%%%%%%%%%%%%%%%
%%%%%%%%%%%%%%%%%%%%%%%%%%%%%%%%%%%%%%%%%%%%%%%%%%%%%%%%%%%%%%%%%%%%%%%%%

This enumeration of models is certainly not exhaustive. There are a number of
single-field models that do not fit well into this scheme, for example
logarithmic potentials $V\left(\phi\right) =V_0\left[1+(C g^2/8\pi)
\ln\left(\phi/\mu\right)\right]$ typical of supersymmetry
\cite{lrreview}, where $C$ counts the degrees of freedom coupled
to the inflaton field and $g$ is a coupling constant.  For this kind of
potential, one gets $n-1\simeq -(1/N)(1+3C g^2/16 \pi^2)$ and $r\simeq (C
g^2/\pi^2)(1/N)$. This model requires an auxiliary field to end inflation
and is more properly categorized as a hybrid model, but falls into the
small-field region of the $r$---$n$ plane.

%%%%%%%%%%%%%%%%%%%%%%%%%%%%%%%%%%%%%%%%%%%%%%%%%%%%%%%%%%%%%%%%%%%%%%%%%
%%%%%%%%%%%%%%%%%%%%%%%%%%%%%%%%%%%%%%%%%%%%%%%%%%%%%%%%%%%%%%%%%%%%%%%%%
\subsection{Beyond first order}
%%%%%%%%%%%%%%%%%%%%%%%%%%%%%%%%%%%%%%%%%%%%%%%%%%%%%%%%%%%%%%%%%%%%%%%%%
%%%%%%%%%%%%%%%%%%%%%%%%%%%%%%%%%%%%%%%%%%%%%%%%%%%%%%%%%%%%%%%%%%%%%%%%%

The four classes of inflation models, categorized by the relationship between
the slow-roll parameters as $-\epsilon < \eta \leq \epsilon$ (large field),
$\eta \leq -\epsilon$ (small field, linear), and $0 < \epsilon < \eta$
(hybrid), cover the entire $r$---$n$ plane and are in that sense complete at
first order in the slow roll parameters. However, this feature is lost going
beyond first order: models can evolve from one region to another.  This feature
is manifest when changing the parameter $N$, and is particularly relevant for
those models for which the running of the observables with the scale is
sizable \cite{kinneyriotto05}. Therefore, it is important to realize that the
lowest-order correspondence between the slow-roll parameters and the class of
models does not always survive to higher order in slow roll.  For instance, for
potentials of the form $ V\left(\phi\right) = \Lambda^4 f\left(\phi/
\mu\right)$, the parameter $\Lambda$ is generally fixed by CMB normalization,
leaving the mass scale $\mu$ and the number of {\it e}-folds $N$ as free
parameters. For some choices of potential, for example $V \propto \exp{(\phi /
\mu)}$ or $V \propto 1 - (\phi / \mu)^2$, the spectral index $n$ varies as a
function of $\mu$. These models therefore appear for fixed $N$ as lines on
$r$---$n$ plane.  Changing $N$ results in a broadening of the lines. For other
choices of potential, for example $V \propto 1 - (\phi / \mu)^p$ with $p > 2$,
the spectral index is independent of $\mu$, and each choice of $p$ describes a
point on the zoo plot for fixed $N$. A change in $N$ turns each of these points
into lines, which smear together into a continuous region.

%%%%%%%%%%%%%%%%%%%%%%%%%%%%%%%%%%%%%%%%%%%%%%%%%%%%%%%%%%%%%%%%%%%%%%%%%
%%%%%%%%%%%%%%%%%%%%%%%%%%%%%%%%%%%%%%%%%%%%%%%%%%%%%%%%%%%%%%%%%%%%%%%%%
\section{Analysis and Results}
\label{secCMBanalysis}
%%%%%%%%%%%%%%%%%%%%%%%%%%%%%%%%%%%%%%%%%%%%%%%%%%%%%%%%%%%%%%%%%%%%%%%%%
%%%%%%%%%%%%%%%%%%%%%%%%%%%%%%%%%%%%%%%%%%%%%%%%%%%%%%%%%%%%%%%%%%%%%%%%%

The method we adopt is based on the publicly available Markov Chain Monte Carlo
(MCMC) package \texttt{cosmomc} \cite{Lewis:2002ah}. We sample the following
eight-dimensional set of cosmological parameters, adopting flat priors on them:
the physical baryon and CDM densities, $\omega_b=\Omega_bh^2$ and
$\omega_c=\Omega_ch^2$, the ratio of the sound horizon to the angular diameter
distance at decoupling, $\theta_s$, the scalar spectral index, its running and
the overall normalization of the spectrum, $n$, $dn/d{\rm ln}\,k$ and $A$ at
$k=0.002$ Mpc$^{-1}$, the tensor contribution $r$, and, finally, the optical
depth to reionization, $\tau$. Furthermore, we consider purely adiabatic
initial conditions, we impose flatness, and we use the inflation consistency
relation to fix the value of the tensor spectral index $n_T$. We also restrict
our analysis to the case of three massless neutrino families; introducing a
neutrino mass in agreement with current neutrino oscillation data doesn't
change our results in a significant way.

We include the three-year data \cite{wmap3cosm} (temperature and polarization)
with the routine for computing the likelihood supplied by the WMAP team and
available at the \texttt{LAMBDA} web
site.\footnote{http://lambda.gsfc.nasa.gov/} We marginalize over the amplitude
of the Sunyaev-Zel'dovich signal, but the effect is small: including/excluding
the correction changes our conclusions on the best fit value of any single
parameter by less than 2\%, and always well within the 68\% C.L.\ contours. We
treat beam errors with the highest possible accuracy (see Ref.\
\cite{wmap3temp}, Appendix A.2), using full off-diagonal temperature covariance
matrix, Gaussian plus lognormal likelihood, and fixed fiducial $C_{\ell}$'s.
The MCMC convergence diagnostic is done on $8$ chains though the Gelman and
Rubin ``variance of chain mean''$/$``mean of chain variances'' $R$ statistic
for each parameter. Our $1-D$ and $2-D$ constraints are obtained after
marginalization over the remaining ``nuisance'' parameters, again using the
programs included in the \texttt{cosmomc} package. In addition to the CMB data,
we also consider the constraints on the real-space power spectrum of galaxies
from the Sloan Digital Sky Survey (SDSS) \cite{thx}. We restrict the analysis
to a range of scales over which the fluctuations are assumed to be in the
linear regime ($k < 0.2 h^{-1}$\rm Mpc). When combining the matter power
spectrum with CMB data, we marginalize over a bias $b$ considered as an
additional nuisance parameter. Furthermore, we make use of the HST measurement
of the Hubble parameter $H_0 = 100h \text{ km s}^{-1} \text{Mpc}^{-1}$
\cite{hst} by multiplying the likelihood by a Gaussian likelihood function
centered around $h=0.72$ and with a standard deviation $\sigma = 0.08$.
Finally, we include a top-hat prior on the age of the universe: $10 < t_0 < 20$
Gyrs.

%%%%%%%%%%%%%%%%%%%%%%%%%%%%%%%%%%%%%%%%%%%%%%%%%%%%%%%%%%%%%%%%%%%%%%%%%
%%%%%%%%%%%%%%%%%%%%%%%%%%%%%%%%%%%%%%%%%%%%%%%%%%%%%%%%%%%%%%%%%%%%%%%%%
\subsection{Results}
%%%%%%%%%%%%%%%%%%%%%%%%%%%%%%%%%%%%%%%%%%%%%%%%%%%%%%%%%%%%%%%%%%%%%%%%%
%%%%%%%%%%%%%%%%%%%%%%%%%%%%%%%%%%%%%%%%%%%%%%%%%%%%%%%%%%%%%%%%%%%%%%%%%

\begin{figure}
\includegraphics[width=3.25in]{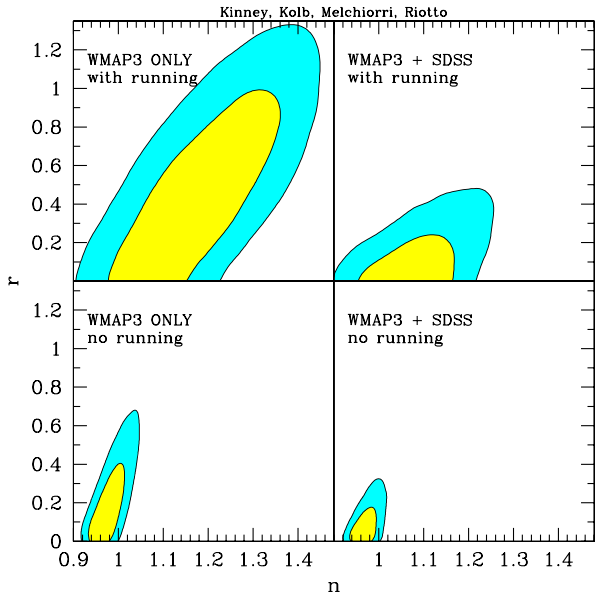}
\caption{\label{fig_rn} Constraints on the
$n$---$r$ plane for different choices of experimental datasets.
The analyses in the top panels include a running spectral index, 
while the analyses in the bottom panels are without running. The shaded 
regions indicate 68\% and 95\% C.L.}
\end{figure}

\begin{figure}
\includegraphics[width=3.25in]{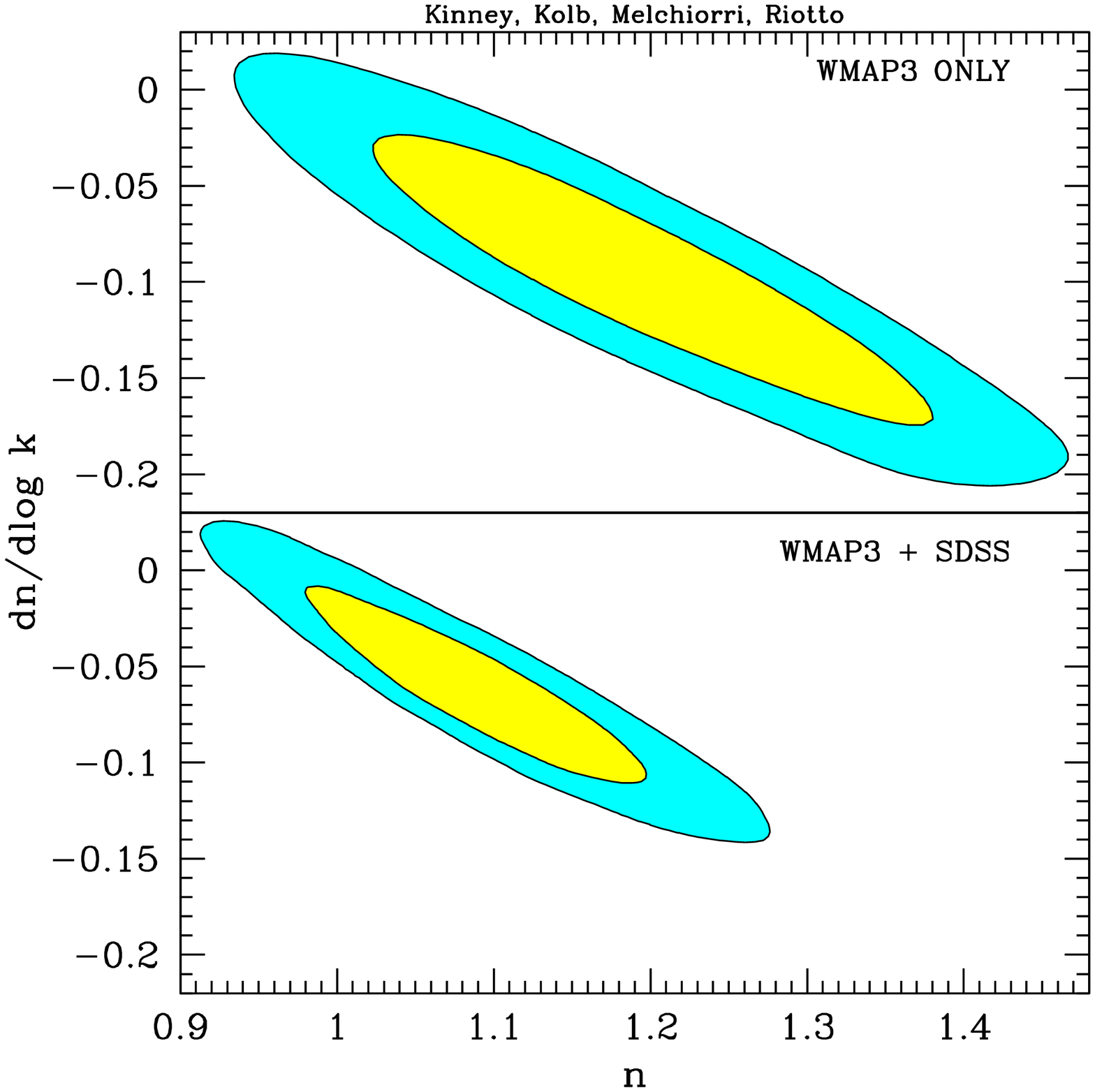}
\caption{\label{fig_dnn} Constraints on the
$n$---$dn/d\ln k$ plane for different choices of experimental datasets.
The shaded regions indicate 68\% and 95\% C.L.}
%\end{figure}
%%%%%%%%%%%%%%%%%%%%%%%%%%%%%%%%%%%%%%%%%%%%%%%%%%%%%%%%%%%%%%%%%%%%%%%%%
%%%%%%%%%%%%%%%%%%%%%%%%%%%%%%%%%%%%%%%%%%%%%%%%%%%%%%%%%%%%%%%%%%%%%%%%%

%%%%%%%%%%%%%%%%%%%%%%%%%%%%%%%%%%%%%%%%%%%%%%%%%%%%%%%%%%%%%%%%%%%%%%%%%
%%%%%%%%%%%%%%%%%%%%%%%%%%%%%%%%%%%%%%%%%%%%%%%%%%%%%%%%%%%%%%%%%%%%%%%%%
%\begin{figure}
\includegraphics[width=3.25in]{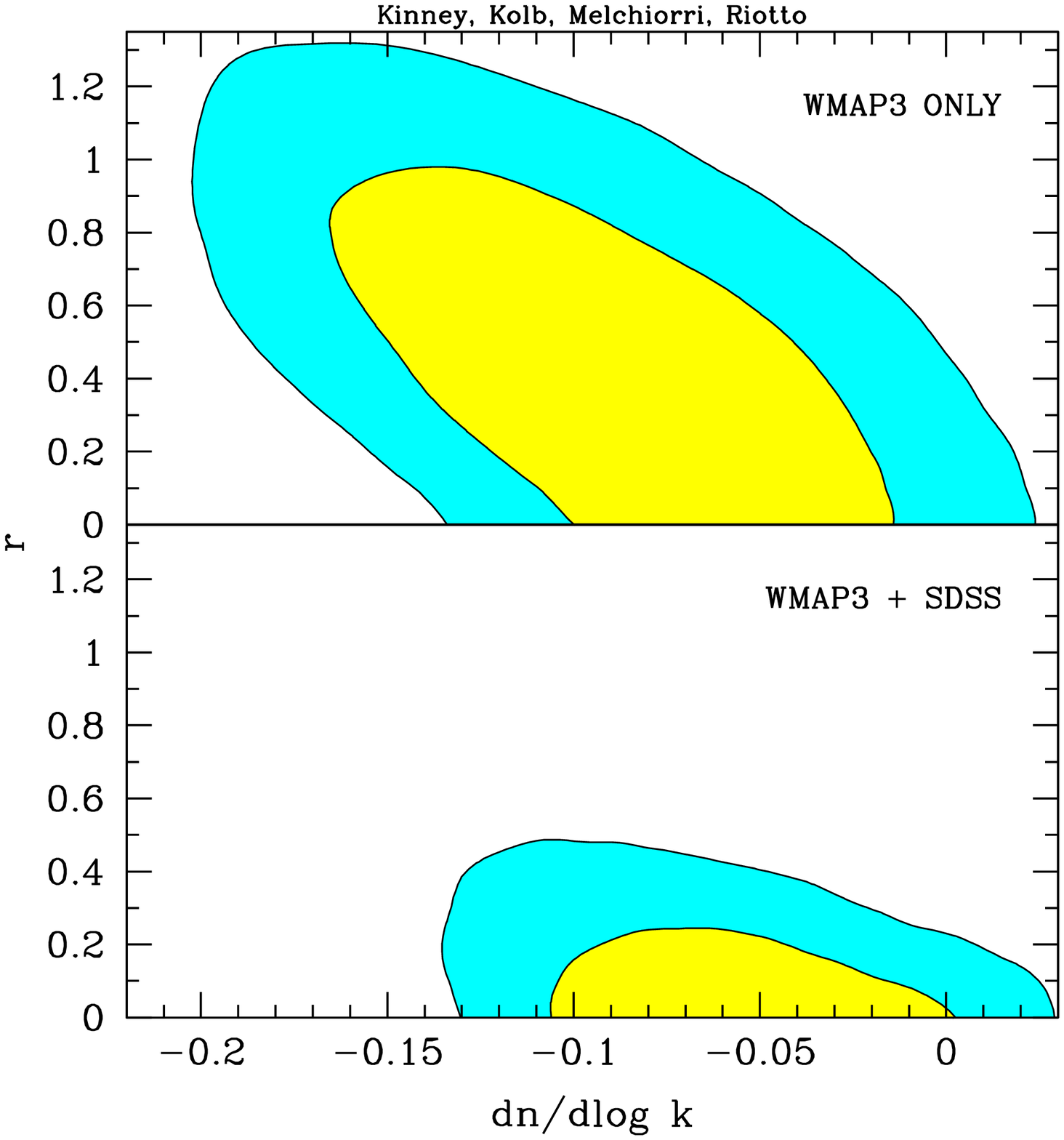}
\caption{\label{fig_rdn} Constraints on the
$dn/d\ln k$---$r$ plane for different choices of experimental datasets.
The shaded regions indicate 68\% and 95\% C.L.}
\end{figure}

As now common practice, we plot the likelihood contours obtained from our
analysis on three different planes, $r$ vs.\ $n$, $dn/d\ln{k}$ vs.\ $n$,  and
$r$ vs.\ $dn/d\ln{k}$: we do so in Figs.\ \ref{fig_rn}--\ref{fig_rdn}. 
Presenting our results on these planes is useful for understanding the effects
of theoretical assumptions and/or external priors.

We consider two different choices of datasets: the WMAP3 dataset alone, and
WMAP3 plus the additional information of the SDSS. By analyzing these different
datasets we can check the consistency of the SDSS large-scale structure data
with WMAP3, something that is not completely trivial since the WMAP3 data seems
to prefer models with a lower value for the $\sigma_8$ parameter than the one
inferred from the SDSS data (see Refs.\ \cite{wmap3temp} and \cite{thx}).

In Fig.\ \ref{fig_rn}, we show the 68\% and 95\% likelihood contours on the $r$
vs.\ $n$ plane in the case of WMAP3 only (left panel) and WMAP3+SDSS (right
panel).  We also consider a prior on the running: the results on the top panel
are obtained allowing the possibility of $dn/d\ln k \neq 0$ while on the bottom
panel assume no running.  The WMAP3 only case including running exhibited 
relatively poor convergence due to a degeneracy in the four-dimensional
parameter  space of $r$, $n$, $dn/d\ln{k}$, and normalization. Adding the SDSS
data set removed the degeneracy and substantially improved the convergence of
the MCMC code. 

Let's first investigate the case of {\it no} running. Marginalizing over all
the remaining nuisance parameters we constrain $n$ and $r$ to $0.94 < n < 1.04$
and $r<0.60$ at $95$\% C.L. Models with $n<0.9$ are  therefore ruled out at
high significance, as are models with $n > 1.05$. The data clearly set
interesting constraints on tensor modes. Models with $n<1$ must have $r<0.4$ at
$95$\% C.L. Models with $n<0.9$ must have a negligible tensor component.
Including the SDSS data further reduces the limit on the amplitude of the
gravitational wave component with a relatively smaller effect on the spectral
index parameter. For WMAP3+SDSS we constrain $n$ and $r$ to $0.93<n<1.01$ and
$r<0.31$.

%%%%%%%%%%%%%%%%%%%%%%%%%%%%%%%%%%%%%%%%%%%%%%%%%%%%%%%%%%%%%%%%%%%%%%%%%
%%%%%%%%%%%%%%%%%%%%%%%%%%%%%%%%%%%%%%%%%%%%%%%%%%%%%%%%%%%%%%%%%%%%%%%%%
%%%%%%%%%%%%%%%%%%%%%%%%%%%%%%%%%%%%%%%%%%%%%%%%%%%%%%%%%%%%%%%%%%%%%%%%%
%%%%%%%%%%%%%%%%%%%%%%%%%%%%%%%%%%%%%%%%%%%%%%%%%%%%%%%%%%%%%%%%%%%%%%%%%

If we allow running the main effect is to open the contours
toward higher value of $n$ and $r$. With running, marginalizing 
over all the remaining nuisance parameters, we constrain $n$ and $r$ 
to $1.02 < n < 1.38$ and $r < 1.09$ at $95$\% C.L.\ for WMAP3 alone
and $0.97 < n < 1.21$ and $r<0.38$ in the case of WMAP3 plus SDSS.

%%%%%%%%%%%%%%%%%%%%%%%%%%%%%%%%%%%%%%%%%%%%%%%%%%%%%%%%%%%%%%%%%%%%%%%%%
%%%%%%%%%%%%%%%%%%%%%%%%%%%%%%%%%%%%%%%%%%%%%%%%%%%%%%%%%%%%%%%%%%%%%%%%%
\begin{table}
\caption{One-dimensional confidence limits on inflationary parameters, 
marginalized over all other parameters, for WMAP3 alone and WMAP3 + SDSS.}
\begin{ruledtabular}
\begin{tabular}{l|c|r}
no running/ & limits on $n$, $r$ & data\\
running & 95\% C.L. & set \\
\hline
& & \\
& $0.94<n<1.04$ & WMAP3 ONLY \\
& $r<0.60$ & WMAP3 ONLY \\
no running & & \\
& $0.93 < n < 1.01$ & WMAP3 + SDSS \\
& $ r<0.31$ & WMAP3 + SDSS \\
& & \\
\hline
& & \\
& $1.02 < n <1.38$ &  WMAP3 ONLY \\
& $ r < 1.09 $ & WMAP3 ONLY \\
& $ -0.17 < dn/d\ln k < -0.02$ & WMAP3 ONLY\\ 
running & & \\
& $0.97 < n < 1.21 $ & WMAP3 + SDSS \\
& $r < 0.38 $ & WMAP3 + SDSS \\
& $-0.13 < dn/d\ln k < 0.007 $ & WMAP3 + SDSS\\ 
& & \\
\end{tabular}
\end{ruledtabular}
\end{table}

%%%%%%%%%%%%%%%%%%%%%%%%%%%%%%%%%%%%%%%%%%%%%%%%%%%%%%%%%%%%%%%%%%%%%%%%%
%%%%%%%%%%%%%%%%%%%%%%%%%%%%%%%%%%%%%%%%%%%%%%%%%%%%%%%%%%%%%%%%%%%%%%%%%

Models with $n=1$ are therefore in very good agreement with CMB data in the
presence of a tensor component or running different from zero. Of particular
interest is the Harrison--Zel'dovich (HZ) model: $n=1$, $r=0$, $dn/d\ln k =0$.
As we see from the bottom panel of  Fig.\ \ref{fig_rn}, pure HZ spectra are not
ruled out at more than 95\% C.L. from CMB data alone. In particular, we found
that, considering the whole sets of models in our $8$-D chain, the HZ best-fit
model is at $\Delta\chi^2/2=2.04$, $2.77$, and $3.96$ with respect to the the
best fit in the case of no running and no tensors,  including tensors but no
running and including tensors and running.  When we include the SDSS data we
found that the HZ best fit model is at $\Delta\chi^2/2=3.07$  with respect to
the best fit in the case of no running and no tensor, $\Delta\chi^2/2=3.4$ with
respect to the best fit with no running and $\Delta\chi^2/2=5.1$  with respect
to the overall best fit.  Since $\Delta\chi^2/2=6.4$ at $95.4$\% confidence
level for $6$ degrees of freedom, those numbers clearly indicate that even when
the SDSS data is included, the HZ spectrum is in reasonable agreement with the
data.

The fact that the scale-invariant value $n=1$ is consistent with the data at
the 95\% C.L.\ when no running is imposed, considerably weakens the bounds
on inflationary models found in Ref.\ \cite{alabidi} where the original WMAP3
error bars were adopted concluding that $n=1$ was ruled out at more than
99\% C.L. 

%%%%%%%%%%%%%%%%%%%%%%%%%%%%%%%%%%%%%%%%%%%%%%%%%%%%%%%%%%%%%%%%%%%%%%%%%
%%%%%%%%%%%%%%%%%%%%%%%%%%%%%%%%%%%%%%%%%%%%%%%%%%%%%%%%%%%%%%%%%%%%%%%%%
%%%%%%%%%%%%%%%%%%%%%%%%%%%%%%%%%%%%%%%%%%%%%%%%%%%%%%%%%%%%%%%%%%%%%%%%%
%%%%%%%%%%%%%%%%%%%%%%%%%%%%%%%%%%%%%%%%%%%%%%%%%%%%%%%%%%%%%%%%%%%%%%%%%

In Fig.\ \ref{fig_dnn} and Fig.\ \ref{fig_rdn} a degeneracy is evident: an
increase in the spectral index $n$ is equivalent to a negative scale dependence
($dn/d\ln{k} < 0$). We emphasize, however, that this behavior depends strictly
on the position of the pivot scale $k_0$: choosing $k_0=0.05h$ Mpc$^{-1}$ would
change the direction of the degeneracy.  Models with $n \sim1.1$ need a
negative running at more than about the 95\% C.L.\ (about $4 \sigma$ in
the case of WMAP3+SDSS). It is interesting also to note that models with a red
spectral index, $n < 1.0$, are in better agreement with the data with a zero or
positive running (see Fig.\ 1\ref{fig_dnn}), while models with a sizable
gravity wave background need a negative running (see Fig.\ \ref{fig_rdn}). For
the WMAP3 alone case the running is bounded by $-0.02 \gtrsim dn/d\ln{k} \gtrsim
-0.17$ at 95\% C.L.\ ($0.007 \gtrsim dn/d\ln{k} \gtrsim -0.13$ for WMAP3+SDSS). 
We found that the best fit from WMAP3 alone with $dn/d\ln k=0$ is at $\Delta
\chi^2/2=1.2$ ($\Delta \chi^2/2=0.2$ when including SDSS) with respect to the
overall best fit. The current data, therefore, do not suggest the presence
of running at more than 95\% C.L.

\begin{figure} \includegraphics[width=3.25in]{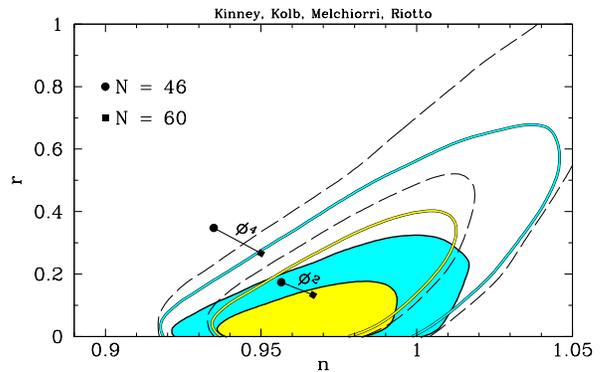}
\caption{\label{fig:zoonorun} The $n$,$r$ parameter space WMAP3 alone (open
contours) and WMAP3 + SDSS (filled contours), with a prior of $dn/d\ln{k} = 0$.
The line segments show the predictions for $V(\phi) = m^2 \phi^2$ and $V(\phi)
= \lambda \phi^4$ for $N = [46,60]$. The dashed lines show the 68\% C.L.\ and
95\% C.L.\ contours from the chains made public by the WMAP team, which do not
include an HST prior on $H_0$ or an age prior. The scale of the plot is chosen
to allow direct comparison with Fig.\ $14$ of Spergel {\it et al.}\
\cite{wmap3cosm}. The shaded regions indicate 68\% and 95\% C.L.} 
\end{figure}

Finally, we compare our results with those presented in Spergel, {\it et al.}\ 
\cite{wmap3cosm}. While there is qualitatively good agreement, a tension
appears when we compare our contour plots in Fig.\ \ref{fig:zoonorun} (the no
running case) with those presented in Fig.\ $14$ of Ref.\ \cite{wmap3cosm}.
Models with a pure HZ spectrum appear to be ruled out by WMAP3 alone at 
about the 99\% C.L.\ in Ref.\ \cite{wmap3cosm}, while our analysis indicates
broader contours, with the Harrison-Zel'dovich spectrum inside the 95\% C.L.
region. Similarly, the contours in Ref.\ \cite{wmap3cosm} appear to rule out
$V(\phi) = \lambda\phi^4$, while our analysis indicates that this potential is
still marginally consistent with the WMAP3 data alone at 95\% confidence.  In
order to better understand this discrepancy, we compared our results directly
with the chains made public by the WMAP team and available at the
\texttt{Lambda} web site.\footnote{\texttt{http://lambda.gsfc.nasa.gov}.}  We
found that the error contours derived from the publicly available chains are
considerably larger than those shown in Fig.\ 14 of Ref.\ \cite{wmap3cosm}:
error contours from our analysis of the WMAP chains are plotted as dashed lines
in Fig.\ \ref{fig:zoonorun}.\footnote{The difference between the WMAP-team
contours and our contours as plotted in Fig.\ \ref{fig:zoonorun} can be
accounted for by the fact that, unlike the WMAP-team analysis, we include
priors on $H_0$ from the HST Key Project data and a top-hat age prior. We have
independently reproduced the dashed-line contours shown in Fig.\
\ref{fig:zoonorun} with our own analysis.}  None of the contours are as tight
as those shown in Spergel {\it et al.}, and the discrepancy is significant
enough to influence important conclusions about the model space, in particular,
the consistency of a Harrison-Zel'dovich spectrum with the data. There appears
to be a clear inconsistency between our results and contours shown in Spergel
{\it et al.}, Figs.\ $12$ and $14$.

%%%%%%%%%%%%%%%%%%%%%%%%%%%%%%%%%%%%%%%%%%%%%%%%%%%%%%%%%%%%%%%%%%%%%%%%%
%%%%%%%%%%%%%%%%%%%%%%%%%%%%%%%%%%%%%%%%%%%%%%%%%%%%%%%%%%%%%%%%%%%%%%%%%
%%%%%%%%%%%%%%%%%%%%%%%%%%%%%%%%%%%%%%%%%%%%%%%%%%%%%%%%%%%%%%%%%%%%%%%%%
%%%%%%%%%%%%%%%%%%%%%%%%%%%%%%%%%%%%%%%%%%%%%%%%%%%%%%%%%%%%%%%%%%%%%%%%%

%%%%%%%%%%%%%%%%%%%%%%%%%%%%%%%%%%%%%%%%%%%%%%%%%%%%%%%%%%%%%%%%%%%%%%%%%
%%%%%%%%%%%%%%%%%%%%%%%%%%%%%%%%%%%%%%%%%%%%%%%%%%%%%%%%%%%%%%%%%%%%%%%%%
\section{Monte Carlo reconstruction}
\label{secmontecarlorecon}
%%%%%%%%%%%%%%%%%%%%%%%%%%%%%%%%%%%%%%%%%%%%%%%%%%%%%%%%%%%%%%%%%%%%%%%%%
%%%%%%%%%%%%%%%%%%%%%%%%%%%%%%%%%%%%%%%%%%%%%%%%%%%%%%%%%%%%%%%%%%%%%%%%%

In this section we describe {\em Monte Carlo reconstruction}, a stochastic
method for ``inverting'' observational constraints to determine an ensemble of
inflationary potentials compatible with observation. The method is described in
more detail in Refs.\ \cite{kinney02,easther02}.  In addition to encompassing a
broader set of models than we considered in the previous section, Monte Carlo
reconstruction allows us easily to incorporate constraints on the running of
the spectral index $d n / d \ln{k}$ as well as to include effects to higher
order in slow roll.

We have defined the slow-roll parameters $\epsilon$ and $\eta$ in terms of
the Hubble parameter $H\left(\phi\right)$ in a previous section.
Taking higher derivatives
of $H$ with respect to the field, we can define an infinite hierarchy of slow
roll parameters \cite{liddle94}:
\begin{eqnarray}
\sigma &\equiv& {m_{\rm Pl} \over \pi} \left[{1 \over 2} \left({H'' \over
 H}\right) -
\left({H' \over H}\right)^2\right],\cr
{}^\ell\lambda_{\rm H} &\equiv& \left({m_{\rm Pl}^2 \over 4 \pi}\right)^\ell
{\left(H'\right)^{\ell-1} \over H^\ell} {d^{(\ell+1)} H \over d\phi^{(\ell +
1)}}.
\end{eqnarray}
Here we have chosen the parameter $\sigma \equiv 2 \eta - 4 \epsilon \simeq n
-1 $ to make comparison with observation convenient.

It is convenient to use $N$ as the measure of time during inflation. As above,
we take $t_e$ and $\phi_e$ to be the time and field value at end of
inflation. Therefore, $N$ is defined as the number of e-folds before the end of
inflation, and increases as one goes {\em backward} in time ($d t > 0
\Rightarrow d N < 0$):
\begin{equation}
{d \over d N} = {d \over d\ln a} = { m_{\rm Pl} \over 2 \sqrt{\pi}}
\sqrt{\epsilon} {d \over d\phi},
\end{equation}
where we have chosen the sign convention that $\sqrt{\epsilon}$ has the same
sign as $H'\left(\phi\right)$:
\begin{equation}
\sqrt{\epsilon} \equiv + {m_{\rm PL} \over 2 \sqrt{\pi}} {H' \over H}.
\end{equation}
Then $\epsilon$ itself can be expressed in terms of $H$ and $N$ simply as
\begin{equation}
\label{eqepsilonfromN}
{1 \over H} {d H \over d N} = \epsilon.
\end{equation}
Similarly, the evolution of the higher-order parameters during inflation is
determined by a set of ``flow'' equations \cite{hoffman00,schwarz01,kinney02},
\begin{eqnarray}
{d \epsilon \over d N} &=& \epsilon \left(\sigma + 2
\epsilon\right),\cr {d \sigma \over d N} &=& - 5 \epsilon \sigma - 12
\epsilon^2 + 2 \left({}^2\lambda_{\rm H}\right),\cr {d
\left({}^\ell\lambda_{\rm H}\right) \over d N} &=& \left[
\frac{\ell - 1}{2} \sigma + \left(\ell - 2\right) \epsilon\right]
\left({}^\ell\lambda_{\rm H}\right) + {}^{\ell+1}\lambda_{\rm
H}.\label{eqfullflowequations}
\end{eqnarray}
The derivative of a slow roll parameter at a given order is higher order in
slow roll. 

A boundary condition can be specified at any point in the
inflationary evolution by selecting a set of parameters
$\epsilon,\sigma,{}^2\lambda_{\rm H},\ldots$ for a given value of $N$. This is
sufficient to specify a ``path'' in the inflationary parameter space that
specifies the background evolution of the spacetime. Taken to infinite order,
this set of equations completely specifies the cosmological evolution, up to
the normalization of the Hubble parameter $H$. Furthermore, such a
specification is exact, with no assumption of slow roll necessary. In practice,
we must truncate the expansion at finite order by assuming that the
${}^\ell\lambda_{\rm H}$ are all zero above some fixed value of $\ell$.  We
choose initial values for the parameters at random from the following ranges:
\begin{eqnarray}
N &=& [46,60]\cr
\epsilon &=& \left[0,0.8\right]\cr
\sigma &=& \left[-0.5,0.5\right]\cr
{}^2\lambda_{\rm H} &=& \left[-0.05,0.05\right]\cr
{}^3\lambda_{\rm H} &=& \left[-0.025,0.025\right],\cr
&\cdots&\cr
{}^{M+1}\lambda_{\rm H} &=& 0.\label{eqinitialconditions}
\end{eqnarray}
Here the expansion is truncated to order $M$ by setting ${}^{M+1}\lambda_{\rm
H} = 0$. In this case, we still generate an exact solution of the background
equations, albeit one chosen from a subset of the complete space of
models. This is equivalent to placing constraints on the form of the potential
$V\left(\phi\right)$, but the constraints can be made arbitrarily weak by
evaluating the expansion to higher order. For the purposes of this analysis, we
choose $M = 5$.  The results are not sensitive to either the choice of order
$M$ (as long as it is large enough) or to the specific ranges from which the
initial parameters are chosen. 

Solutions to the truncated flow equations are solutions for which all of the
derivatives of the Hubble constant above order $M + 1$ vanish:
\begin{equation}
{d^\ell H \over d \phi^\ell} = 0,\ \ell \geq M + 2,
\end{equation}
with a simple polynomial solution \cite{Liddle:2003py},
\begin{equation}
H\left(\phi\right) = H_0 \left(1 + A_1 \phi + \cdots + A_{M + 1} \phi^{M +
1}\right).
\end{equation}
The Hamilton-Jacobi Equation (\ref{eqhubblehamiltonjacobi}) can be applied to
this solution to derive an analytic form for the potential in terms of the
parameters $A_1,\ldots,A_{M + 1}$. The set of boundary conditions in Eq.\
(\ref{eqinitialconditions}) then consist of a weak slow-roll prior on the
polynomial fit for $H(\phi)$: the inflaton must be slowly rolling at least at
one point in its evolution. Thus, while the flow equations in and of themselves
simply define an expansion in $H(\phi)$, the choice of boundary condition and
the requirement that inflation last at least 46 e-folds comprise a well-defined
physical prior on the inflationary model space. 

Some interesting recent papers have explored alternative methods for
constraining the ``model space'' of inflation. In particular, Ref.\
\cite{Peiris:2006ug} incorporates the lowest-order flow parameters directly
into the Monte Carlo Markov Chain fit, although they do not include effects to
higher order in slow roll. Refs.\ \cite{Parkinson:2006ku,Pahud:2006kv} apply a
Bayesian model selection approach to the problem, but also do not consider
higher-order effects which in principle contribute to a running spectral index.
Our analysis extends these results by including running of the spectral index
as well as effects to higher order in slow roll. 

Once we obtain a solution to the flow equations
$[\epsilon(N),\sigma(N),{}^\ell\lambda_{\rm H}(N)]$, we can calculate the
predicted values of the tensor/scalar ratio $r$, the spectral index $n$, and
the ``running'' of the spectral index $d n / d\ln k$.  To lowest order, the
relationship between the slow roll parameters and the observables is especially
simple: $r = 16 \epsilon$, $n - 1 = \sigma$, and $d n / d \ln k = 0$. To
second order in slow roll, the observables are given by
\cite{liddle94,stewart93},
\begin{equation}
r = 16 \epsilon \left[1 - C \left(\sigma + 2 \epsilon\right)\right]
\label{eqrsecondorder}
\end{equation}
for the tensor/scalar ratio, and 
\begin{equation}
n - 1 = \sigma - \left(5 - 3 C\right) \epsilon^2 - {1 \over 4} \left(3
- 5 C\right) \sigma \epsilon + {1 \over 2}\left(3 - C\right)
\left({}^2\lambda_{\rm H}\right)
\label{eqnsecondorder}
\end{equation}
for the spectral index. The constant $C \equiv 4 (\ln{2} +
\gamma) - 5 = 0.0814514$, where $\gamma \simeq 0.577$ is Euler's
constant. Derivatives
with respect to wavenumber $k$ can be expressed in terms of derivatives with
respect to $N$ as \cite{liddle95}
\begin{equation}
{d \over d N} = - \left(1 - \epsilon\right) {d \over d \ln k} .
\end{equation}
The scale dependence of $n$ is then  given by the simple expression
\begin{equation}
{d n \over d \ln k} = - \left({1 \over 1 - \epsilon}\right) {d n \over d N},
\end{equation}
which can be evaluated by using Eq.~(\ref{eqnsecondorder}) and the flow
equations.  For example, for the case of $V \propto \phi^4$, 
the observables to lowest order are
\begin{eqnarray}
\label{eqphi4obs}
r &\simeq& {16 \over N + 1},\cr
n - 1 &\simeq& - {3 \over N + 1},\cr
{dn \over d\ln k} &\simeq& - {3 \over N \left(N + 1\right)}.
\end{eqnarray}
The final result following the evaluation of a particular path in the
$M$-dimensional ``slow-roll space'' is a point in ``observable parameter
space,'' {\em i.e.,} $(r,n,dn/d\ln k)$, corresponding to the 
observational prediction
for that particular model. 

The reconstruction method works as follows:
\begin{enumerate}
\item Specify a ``window'' of parameter space: {\em e.g.,} central values for
$n-1$, $r$, or $d n /d \ln{k}$ and their associated error bars.
\item Select a random point in slow roll space, 
$[\epsilon,\eta,{}^\ell\lambda_{\rm H}]$, truncated at order $M$ in
the slow roll expansion.
\item Evolve forward in time ($d N < 0$) until either (a) inflation ends
 ($\epsilon > 1$), or (b) the evolution reaches a late-time fixed
 point ($\epsilon = {}^\ell\lambda_{\rm H} = 0,\ \sigma = {\rm
 const.}$).
\item If the evolution reaches a late-time fixed point, calculate the
 observables $r$, $n - 1$, and $d n / d \ln k$ at this point.
\item If inflation ends, evaluate the flow equations backward $N$ e-folds from
 the end of inflation. Calculate the observable parameters at that
 point.
\item If the observable parameters lie within the specified window of
parameter  space, compute the potential and add this model to the ensemble 
of ``reconstructed'' potentials.
\item Repeat steps 2 through 6 until the desired number of models
have been found.
\end{enumerate}

We performed the Monte Carlo reconstruction using the more restrictive of the
data sets considered, combining the WMAP3 data with the Sloan Digital Sky
Survey. We ran the reconstruction code long enough (10,703,502
iterations) to collect 10,000 models consistent with the WMAP3 + SDSS error
bars: 4115 are within the 68\% C.L.\ contours, and 5885 are within the 
95\% C.L.\ contours. 

\begin{figure}
\includegraphics[width=3.25in]{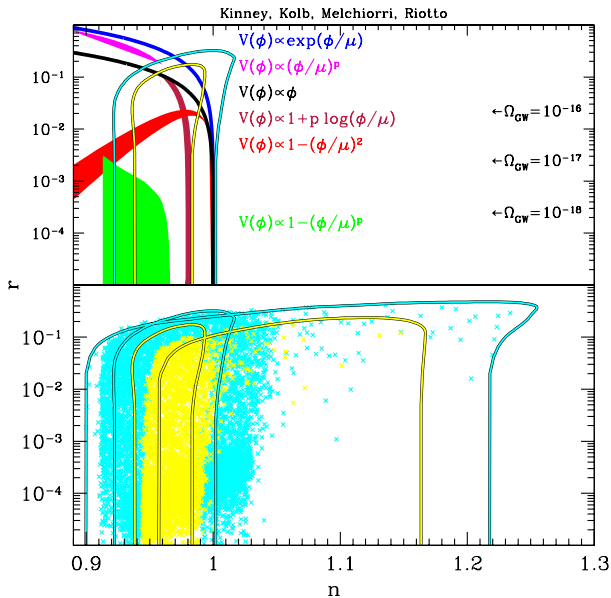}
\caption{\label{fig:MCR_log} Inflationary models plotted against the 
68\% and 95\% WMAP3 +
SDSS error contours. Top panel: the predictions of various specific
inflationary potentials (solid bands) plotted against the error bars from WMAP3
+ SDSS with a prior of $dn / d\ln{k} = 0$. Bottom panel: 10,000 models
generated by flow Monte Carlo consistent with the WMAP3 + SDSS data sets
including running as a parameter, indicated by the larger error contours. The
contours with a $dn / d\ln{k} = 0$ prior are plotted as a reference, and were
not used in the Monte Carlo reconstruction. (Some data points fall outside the
error contours plotted because likelihoods for the models were calculated using
the full three-dimensional likelihood function ${\cal L}(n,r,dn/d\ln{k})$, and
the contours were obtained by marginalizing over $dn/d\ln{k}$).}
\end{figure}

To illustrate the degree of overlap between the various classes of model, the
predictions for different models are shown in the top panel of Fig.\
\ref{fig:MCR_log}, including the effect of the uncertainty in the number of
e-folds $N$. The different classes of potential do not have significant
overlap,  and it is therefore possible to distinguish one  from another
observationally.

Figure \ref{fig:MCR_log} also shows the points generated by Monte Carlo
Reconstruction in the $n,r$ parameter space. Since there is no measure on the
space of initial conditions, the distribution of points generated by the flow
equations cannot be interpreted in a rigorously statistical fashion: the error
bars are those generated from the data using COSMOMC, and the points plotted
are points generated by the flow equations consistent with those errors,
including running of the spectral index as a parameter. The clustering of the
models in the parameter space, however, {\em is} significant: selecting models
based on even an extremely weak assumption of slow-roll results in a strong
clustering of the models in the region favoring a red spectrum and $dn/d\ln{k}
= 0$. 

In this sense, the preference for running and a blue spectrum present in the
data itself contains very little information relevant to constraining slow-roll
inflation models: it can be interpreted simply an artifact of an ``accidental''
parameter degeneracy in the data. Allowing running as a parameter but assuming
slow roll inflation gives constraints on the inflationary model space largely
consistent with an analysis which assumes negligible running as a prior on the
parameter space from the beginning. In  other words: {\em there is no evidence
for inflation with a measurable running of the spectral index.}

%%%%%%%%%%%%%%%%%%%%%%%%%%%%%%%%%%%%%%%%%%%%%%%%%%%%%%%%%%%%%%%%%%%%%%%%%
%%%%%%%%%%%%%%%%%%%%%%%%%%%%%%%%%%%%%%%%%%%%%%%%%%%%%%%%%%%%%%%%%%%%%%%%%
%%%%%%%%%%%%%%%%%%%%%%%%%%%%%%%%%%%%%%%%%%%%%%%%%%%%%%%%%%%%%%%%%%%%%%%%%
%%%%%%%%%%%%%%%%%%%%%%%%%%%%%%%%%%%%%%%%%%%%%%%%%%%%%%%%%%%%%%%%%%%%%%%%%

{}From the flow equations (\ref{eqfullflowequations}) it is evident that the
line along the $r= 0$ axis, with $\epsilon = {}^\ell\lambda_{\rm H} = 0$ is a
fixed point of the flow evolution, including taking the flow equations to
infinite order.\footnote{See Refs.\ \cite{kinney02,Chongchitnan:2005pf} for a
detailed discussion of the fixed-point structure of the slow roll space.}  For
parameters on the ``red'' side of scale invariance, {\it i.e.} $\sigma < 0$,
this fixed point is {\em unstable}: flow moves away from the fixed point as $N
\rightarrow 0$, and toward the fixed point as $N \rightarrow \infty$.
Conversely, the fixed point for $\sigma > 0$ is {\em stable}: models evolve
toward this fixed point at late times, $N \rightarrow 0$.  Integrating the flow
equations forward in time yields one of two possible outcomes. One possibility
is that the condition $\epsilon = 1$ may be satisfied for some finite value of
$N$, which defines the end of inflation. We identify this point as $N=0$ so
that the primordial fluctuations are actually generated when $N = [46,60]$.
Alternatively, the solution can evolve toward an inflationary fixed point with
$r = 0$ and $n > 1$, in which case inflation never stops. In reality, inflation
must stop at some point, presumably via some sort of instability, such as the
``hybrid'' inflation mechanism \cite{linde91,linde94,copeland94,lr97}. Examples
of potentials which fall into this class of models are the simplest hybrid
potentials,
\begin{equation}
V\left(\phi\right) = \Lambda^4 \left[1 + \left({\phi \over
\mu}\right)^p\right].
\end{equation}

Here we take the observables for such models to be the values at the late-time
attractor. Since models on the attractor are by definition those for which the
variation in the slow roll parameters with $N$ vanishes, such models also
predict zero running of the scalar spectral index.  We find that {\em the WMAP3
data strongly disfavor models which evolve to a late-time asymptote with $r =
0$, $n > 1$, and $dn / d \ln{k} = 0$.} The $95$\% confidence limit for a blue
spectrum with no tensors and no running ({\it i.e.}, not marginalized over $r$)
from WMAP3 alone is $n < 1.0007$, and from WMAP3 + SDSS is $n < 1.001$.  Of
more  than  ten million models tested, only one model consistent with the data
relaxed  to the late-time asymptote, with a spectral index $n = 1.0004$ and $r
= 0.0000002$; for all intents and purposes a Harrison-Zel'dovich spectrum.
Every other model in the Monte Carlo reconstruction set was of the
``nontrivial'' type, with inflation ending naturally by evolving through
$\epsilon = 1$ at late times. We note that the level of running required to
accommodate a blue spectrum is severe:  even dynamical supersymmetric
inflation, which predicts a blue spectrum and negative running [Eq.\
(\ref{eq:dsirunning})], does not produce a strong enough running to  match the
data, and is also ruled out to more than 95\% C.L.\ by WMAP3 + SDSS for  $n >
1.001$.

%%%%%%%%%%%%%%%%%%%%%%%%%%%%%%%%%%%%%%%%%%%%%%%%%%%%%%%%%%%%%%%%%%%%%%%%%
%%%%%%%%%%%%%%%%%%%%%%%%%%%%%%%%%%%%%%%%%%%%%%%%%%%%%%%%%%%%%%%%%%%%%%%%%
\begin{figure}
\includegraphics[width=3.25in]{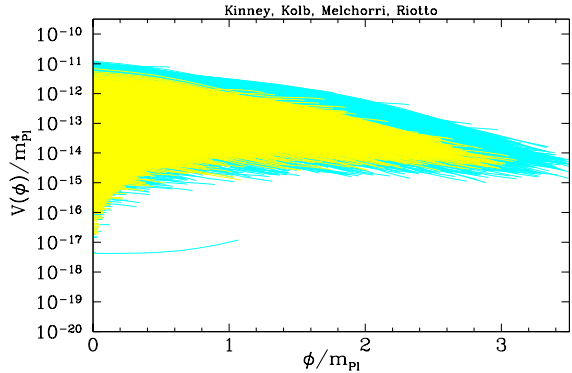}
\caption{\label{fig:V_recon} Potentials generated by Monte Carlo Reconstruction
consistent with WMAP3 + SDSS to 68\% C.L.\ in the light shaded (yellow) region 
and 95\% C.L.\ in the darker shaded (cyan) region. The WMAP3 data place an
upper limit of about $2 \times 10^{16}\ {\rm GeV}$ on the energy scale of
inflation. No lower limit is possible without a detection of a tensor mode
signal. The concave-up line at the bottom of the figure is the single model in
$10^{7}$ models generated which converged to an inflationary fixed point at
late time.}
\end{figure}
%%%%%%%%%%%%%%%%%%%%%%%%%%%%%%%%%%%%%%%%%%%%%%%%%%%%%%%%%%%%%%%%%%%%%%%%%
%%%%%%%%%%%%%%%%%%%%%%%%%%%%%%%%%%%%%%%%%%%%%%%%%%%%%%%%%%%%%%%%%%%%%%%%%

We can also place constraints on the energy scales relevant to inflation, in
particular the ``height'' of the potential $V \sim \Lambda^4$, and the
``width'' of the potential, typically quantified as the field variation $\Delta
\phi$ during inflation. Given a path in the slow roll parameter space, the
form of the potential is fixed, up to normalization
\cite{hodges90,copeland93,beato00,easther02}. The starting point is the
Hamilton-Jacobi equation,
\begin{equation}
V(\phi) = \left({3 m_{\rm Pl}^2} \over 8 \pi\right)
H^2(\phi) \left[1 - {1\over 3}
\epsilon(\phi)\right].\label{eqHJpotential}
\end{equation}
We have $\epsilon(N)$ trivially from the flow equations. In order to calculate
the potential, we need to determine $H(N)$ and $\phi(N)$. With $\epsilon$
known, $H(N)$ can be determined by inverting the definition of $\epsilon$, Eq.\
(\ref{eqepsilonfromN}).  Similarly, $\phi(N)$ follows from the first
Hamilton-Jacobi equation, Eq.\ (\ref{eqbasichjequations}):
\begin{equation}
{d \phi \over d N} = {m_{\rm PL} \over 2 \sqrt{\pi}}
\sqrt{\epsilon}.
\end{equation}
Using these equations and Eq.~(\ref{eqHJpotential}), the form of the potential
can then be fully reconstructed from the numerical solution for $\epsilon(N)$.
The only necessary observational input is the normalization of the Hubble
parameter $H$, which enters the above equations as an integration constant.
Here we use the simple condition that the density fluctuation amplitude (as
determined by a first-order slow roll expression) be of order $10^{-5}$,
\begin{equation}
{\delta \rho \over \rho} \simeq {H \over m_{\rm Pl}} 
\frac{1}{\sqrt{\pi \epsilon}} = 10^{-5}.
\end{equation}
A more sophisticated treatment would perform a full normalization to the
CMB data \cite{bunn94,stompor95}.  The value of the field, $\phi$, also
contains an arbitrary, additive constant. Fig.\ \ref{fig:V_recon} shows the
reconstructed potentials consistent with the WMAP3 + SDSS data set. 

We see that the energy scale for inflation favored by the flow Monte Carlo
gives $V_0$ between $5 \times 10^{14}\ {\rm GeV}$ and $2 \times 10^{16}\ {\rm
GeV}$, although without a detection of a nonzero tensor/scalar ratio, it is not
possible to put a purely observational lower limit on the height of the
inflationary potential. Inflationary potentials with very low energy scales (in
particular those with $\Delta \phi \ll m_{\rm Pl}$ during inflation) require
the imposition of a symmetry to suppress the mass term for the inflaton
\cite{Knox:1992iy,Kinney:1995cc,Easther:2006qu}. Such potentials (corresponding
to the region labeled $V(\phi) \propto 1 - (\phi / \mu)^p$ in Fig.\
\ref{fig:MCR_log}) are consistent with the WMAP3 data and predict an
unobservably small value for the tensor/scalar ratio $r$. Even if one wished to
consider such models fine-tuned (see, {\it e.g.} Ref.\
\cite{Efstathiou:2006ak}) and therefore disfavored, the range of energy scales
favored by the flow Monte Carlo (Fig.\ \ref{fig:V_recon}) is consistent with
tensor/scalar ratios as low as $r \sim 10^{-5}$, a level unlikely to be
detectable by any currently foreseen experiments. Refs.\
\cite{Boyle:2005ug,Bock:2006yf} suggest that fine-tuning considerations force
the tensor/scalar ratio for a red spectrum to detectable levels of $r \sim
0.01$. We see no evidence of such an effect in the flow analysis, for which no
explicit tuning of the potential is performed.

%%%%%%%%%%%%%%%%%%%%%%%%%%%%%%%%%%%%%%%%%%%%%%%%%%%%%%%%%%%%%%%%%%%%%%%%%
%%%%%%%%%%%%%%%%%%%%%%%%%%%%%%%%%%%%%%%%%%%%%%%%%%%%%%%%%%%%%%%%%%%%%%%%%
\section{Conclusions}
%%%%%%%%%%%%%%%%%%%%%%%%%%%%%%%%%%%%%%%%%%%%%%%%%%%%%%%%%%%%%%%%%%%%%%%%%
%%%%%%%%%%%%%%%%%%%%%%%%%%%%%%%%%%%%%%%%%%%%%%%%%%%%%%%%%%%%%%%%%%%%%%%%%

In this paper, we presented an analysis of the recent WMAP three-year data set
with an emphasis on parameters relevant for distinguishing among the various
possible models for inflation. Our results are in good agreement with other
analyses of the data \cite{Lewis:2006ma,Seljak:2006bg}, but show significant
inconsistencies with the results reported by the WMAP team in Figs.\ $12$ and
$14$ of Ref.\ \cite{wmap3cosm}.

We found that the WMAP3 data alone are consistent within 95\% C.L.\ with a
scale-invariant power spectrum, $n = 1$, with no running of the spectral index,
$dn/d\ln{k} = 0$ and no tensor component. The Harrison-Zel'dovich spectrum is
therefore still not ruled out at high significance, a conclusion in accord with
Refs.\ \cite{Parkinson:2006ku,Magueijo:2006we}.  While a detection of a running
spectral index would be of great significance for inflationary model building
\cite{Easther:2006tv,Cline:2006db}, no clear evidence for  running is present
in the WMAP3 data. The data are, however, consistent with strongly negative
running combined with a large tensor/scalar ratio and a ``blue'' power
spectrum. 

The inclusion of the Sloan Digital Sky Survey datasets in the analysis has the
effect  of reducing the error bars and gives a better determination of the
inflationary parameters. For instance, the inclusion of SDSS rules out  quartic
chaotic models of inflation of the form $V(\phi)\sim \lambda \phi^4$. Chaotic
inflation with a quadratic potential  $V(\phi) \sim m^2 \phi^2$ is consistent
with all datasets considered. 

In addition, we applied the Monte Carlo reconstruction technique to generate an
ensemble of inflationary potentials consistent with observation.  Our results
may be summarized as follows: Models which evolve to a late-time fixed point in
the space of flow parameters are strongly disfavored by the data. Of more than
10 million models analyzed in the flow  Monte Carlo, one evolved to a late-time
inflationary asymptote indistinguishable from a Harrison-Zel'dovich spectrum.
The rest were characterized by a dynamical end to inflation, with the first
slow-roll parameter $\epsilon$ evolving to unity in finite time. The late-time
attractor in flow space corresponds to models with a ``blue'' power spectrum
($n > 1$) and negligible $r$ and $dn/d\ln{k}$, and we conclude that such
models are inconsistent with the data for $n > 1.001$.  This is a significant
constraint on the inflationary model space. In particular, the  data  rule out
the simplest models of hybrid inflation of the form $V(\phi) = V_0 + m^2
\phi^2$ as well as models such as $V(\phi) \propto 1 + (\mu / \phi)^{p}$, which
predict some negative running. This of course does not rule out all models for
which inflation ends via a hybrid mechanism.  Some  hybrid models are
characterized by a red spectrum, for example ``inverted'' hybrid models and
models with logarithmic potentials inspired by (global) supersymmetry. Finally,
we find that there is no evidence to support any  lower bound on the amplitude
of gravitational waves. Tensor/scalar ratios as low as $r \sim 10^{-5}$ were
produced by the flow Monte Carlo without explicit tuning of the inflationary
potential.

%%%%%%%%%%%%%%%%%%%%%%%%%%%%%%%%%%%%%%%%%%%%%%%%%%%%%%%%%%%%%%%%%%%%%%%%%
%%%%%%%%%%%%%%%%%%%%%%%%%%%%%%%%%%%%%%%%%%%%%%%%%%%%%%%%%%%%%%%%%%%%%%%%%
\acknowledgments
%%%%%%%%%%%%%%%%%%%%%%%%%%%%%%%%%%%%%%%%%%%%%%%%%%%%%%%%%%%%%%%%%%%%%%%%%
%%%%%%%%%%%%%%%%%%%%%%%%%%%%%%%%%%%%%%%%%%%%%%%%%%%%%%%%%%%%%%%%%%%%%%%%%

We thank Rachel Bean, Olivier Dore, Richard Easther, Justin Khoury, Hiranya
Peiris, and Licia Verde for helpful conversations. We acknowledge support
provided by the Center for Computational Research at the University at
Buffalo. WHK is supported in part by the National Science Foundation under
grant NSF-PHY-0456777.  EWK is supported in part by NASA (NAG5-10842).

%%%%%%%%%%%%%%%%%%%%%%%%%%%%%%%%%%%%%%%%%%%%%%%%%%%%%%%%%%%%%%%%%%%%%%%%%
%%%%%%%%%%%%%%%%%%%%%%%%%%%%%%%%%%%%%%%%%%%%%%%%%%%%%%%%%%%%%%%%%%%%%%%%%

%%%%%%%%%%%%%%%%%%%%%%%%%%%%%%%%%%%%%%%%%%%%%%%%%%%%%%%%%%%%%%%%%%%%%%%%%
%%%%%%%%%%%%%%%%%%%%%%%%%%%%%%%%%%%%%%%%%%%%%%%%%%%%%%%%%%%%%%%%%%%%%%%%%

\end{document}